\documentclass[a4paper,12pt]{article}
\usepackage{amssymb}
\usepackage{latexsym}
\usepackage[dvips]{graphicx}
\usepackage{cite}

\usepackage{ae} 
\usepackage[T1]{fontenc}
\usepackage[ansinew]{inputenc}
\usepackage{amsmath}
\usepackage{graphicx}
\usepackage{color}

\begin{document}
\begin{center}
{\LARGE{\bf Geodesics and Geodesic Deviation in\\[.5em] Static Charged Black Holes}}\\[2em]
\large{\bf{Ragab M. Gad}\footnote{Email Address: ragab2gad@hotmail.com}}\\
\normalsize {Mathematics Department, Faculty of Science,}\\
\normalsize  {Minia University, 61915 El-Minia,  EGYPT.}
\end{center}

\begin{abstract}
The radial motion along null geodesics in static charged black hole
space-times, in particular, the Reissner-Nordstr\"om and stringy
charged black holes are studied. We analyzed the properties of the
effective potential. The circular photon orbits in these space-times
are investigated. We found that the radius of circular photon orbits
in both charged black holes are different and  differ from that
given in Schwarzschild space-time. We studied the physical effects of
the gravitational field between  two test particles in stringy
charged black hole and compared the results with that given in
Schwarzschild and Reissner-Nordstr\"om black holes.

\end{abstract}
{\bf{PACS numbers}} : 04.50.+h, 12.10.-g\\
{\bf{Keywords}}: Geodesics and geodesic  deviation;  circular photon orbits; static charged black hole.
\setcounter{equation}{0}

\section{Introduction}
The well known static, spherically symmetric black hole solutions in
vacuum of Einstein's general relativity are  given by the charged
Reissner-Nordstr\"{o}m and uncharged Schwarzschild solutions. In the
non-vacuum case of Einstein's general relativity, several black hole
solutions are known \cite{B75}-\cite{G91}. One of them is the
stringy charged black hole discovered by Garfinkle, Horowitz and
Strominger \cite{G91}.
\par
The study of timelike and null geodesics, the paths of freely moving
particles and photons, is the key to understand the physical
importance of a given space-time.
\par
We wish to investigate in this paper the properties of the stringy
charged black hole by studying its geodesic structure, that is, from
the motion of  photons. We compare the results by the aforementioned
solutions in vacuum.
\par
The space-time under consideration is almost identical to
Schwarzschild space-time. The only  differences are
\begin{enumerate}
\item The gravitational energy, using M{\o}ller's prescription
\cite{M58}, depends on the mass parameter $M$ and on the charge $Q$
(see Gad \cite{G04}), while in the Schwarzschild black hole is given
only by the mass parameter $M$ (see Xulu \cite{X}). \item The areas
of the spheres of constant $r$ and $t$ depend on the charge $Q$.
\end{enumerate}
The first aim of this paper is to sustain the second difference by
analyzing the effective potential of radial motion along null
geodesics.
\par
Compared to the Reissner-Nordstr\"{o}m black hole, the stringy
charged black hole exhibits several different properties
\cite{PSSTW91}-\cite{T83}. For example, first, this solution has
only one horizon at $r=2M$ and not two as is the case for the
Reissner-Nordstr\"{o}m\footnote{The Reissner-Nordstr\"{o}m has two
horizons given by the quadratic equation $r^2 -2Mr +e^2 =0$.}.
Second, this solution has singularities at
$r={\bf{\alpha}}$, $r=0$ whereas the
Reissner-Nordstr\"{o}m has a singularity at $r=0$. The extremal solution occurs at $Q^2=4M^2e^{2\Phi_{0}}$ rather than at $E^2=M^2$ as for Reissner-Nordstr\"{o}m solution.\\
It is therefore worthwhile to investigate other properties of the
stringy charged black hole to see how this differs from the vacuum
solutions.
\par
The equation of geodesic deviation gives the relative accelerations
between free test particles falling in a gravitational field and is
a cornerstone to the understanding of the physical effects of the
gravitational field  and the geometry of space-time \cite{FAE}. Geodesics deviations in vacuum
space-times, namely, Schwarzschild and Reissner-Nordstr\"{o}m
space-times, are rigorously studied in \cite{MTW73}-\cite{AG05}. Some studies on geodesic deviation can be found in \cite{KMMH01}-\cite{KK03}. Ghosh and Kar \cite{GK09} investigated geodesic motion and geodesic deviation in warped space-times with a  a time dependent extra dimension.
We
do follow here the same approach in \cite{D92,AG05} to obtain the
relative accelerations of nearby test particles in stringy charged
black hole. This is the second aim of this paper.
\par
The structure of the paper is as  follows: In the next section, we
investigate the geodesic curves and the circular photon orbits in
the stringy charged black hole. In section 3, we study the photon
trajectories in Reissner-Nordst\"{o}rm by analyzing the properties
of effective potential. Section 4 concerns with the tidal forces
between nearby test particles in stringy charged black hole. In section 5, we investigate the behavior of geodesic deviation vector along a timelike geodesic near the singularities $r=0$ and $r={\bf{\alpha}}$, by considering a special case.
Finally, in section 6, brief summary of results and concluding
remarks are presented.\\
Through out this paper Latin indices run from $0$ to $3$; Greek indices
run from $1$ to $3$. Units are such that $G=1=c$, ($G$ is the
gravitational constant; $c$ is the velocity of light).

 \setcounter{equation}{0}
\section{The equations of geodesic motion}
In recent years there is considerable interest in obtaining black
hole solutions in string theory and investigating their properties
(see references \cite{G91},\cite{PSSTW91} and  \cite{STW91}). There
are two metrics in this theory, which are called the sigma-model (or
string) metric and Einstein metric. For uncharged static,
spherically symmetric black hole, the solution in the low energy is
the same as the Schwarzschild solution. This is,
however, not the case when the black hole is charged.\\

 We are interested to investigate the geodesic curves of a static
spherically symmetric charged black hole. The line element
representing this space-time is given by Garfinkle et al \cite{G91}.
\begin{equation}\label{eq}
ds^2=-(1-\frac{2M}{r})dt^2 + (1-\frac{2M}{r})^{-1} dr^2
+(1-\frac{{\bf{\alpha}}}{r})r^2(d\theta^2 +\sin^2\theta d\phi^2),
\end{equation}
where
\begin{equation}\label{Q}
{\bf{\alpha}} =\frac{Q^2}{2M}exp(-2\Phi_{0}),
\end{equation}
$M$ and $Q$ are, respectively, mass and charge  parameters;
$\Phi_{0}$ is the asymptotic value of dilaton field.
\par
The equations and constraint for geodesics are given as
$$
\ddot{x}^a +\Gamma^a_{bc}\dot{x}^b \dot{x}^c =0,
$$
$$
g_{ab}\dot{x}^a\dot{x}^b = \varepsilon,
$$
where a superposed dot stands for a derivative with respect to the
affine parameter $\tau$ associated to the geodesic, $x^a$ are the
coordinates of a space-time point on the geodesic and
$\varepsilon=-1$ or $0$, for timelike  or null  geodesics,
respectively.
\par
The geodesic equations for the line element (\ref{eq}) are given by
\begin{equation}\label{t}
\ddot{t}+\frac{2M}{r(r-2M)}\dot{r}\dot{t}=0,
\end{equation}
$$
\ddot{r}-\frac{M}{r(r-2M)}\dot{r}^2+\frac{(r-2M)({\bf{\alpha}}
-2r)}{2r}\dot{\theta}^2 +\frac{(r-2M)({\bf{\alpha}}
-2r)}{2r}\sin^2\theta\dot{\phi}^2+
$$
\begin{equation}\label{r}
\frac{M(r-2M)}{r^3}\dot{t}^2=0,
\end{equation}
\begin{equation}\label{theta}
\ddot{\theta} +\frac{({\bf{\alpha}}-2r)}{r({\bf{\alpha}}
-r)}\dot{r}\dot{\theta} -\sin\theta\cos\theta\dot{\phi}^2=0,
\end{equation}
\begin{equation}\label{phi}
\ddot{\phi} +\frac{({\bf{\alpha}}-2r)}{r({\bf{\alpha}}
-r)}\dot{r}\dot{\phi} +2\cot\theta\dot{\theta}\dot{\phi}=0,
\end{equation}

\par
The constraint of timelike or null geodesics for the line element
(\ref{eq}) is given by
\begin{equation}\label{Cr}
-(1-\frac{2M}{r})\dot{t}^2 + (1-\frac{2M}{r})^{-1} \dot{r}^2
+(1-\frac{{\bf{\alpha}}}{r})r^2(\dot{\theta}^2 +\sin^2\theta
\dot{\phi}^2)=\varepsilon.
\end{equation}
\par
Let a geodesic $\gamma$ be given by $\gamma(\tau)=(t(\tau), r(\tau),
\theta(\tau), \phi(\tau))$. We  suppose without loss of
generality that $\gamma(\tau_{0})=(t(\tau_{0}), r(\tau_{0}),
\frac{\pi}{2}, \phi(\tau_{0}))$ and for all $\tau : \theta =
\frac{\pi}{2}$ (equatorial orbits ) that is the particle has at
start, and continues to have, zero momentum in the
$\theta$-direction; thus $\dot{\theta}=0$. Consequently, the
equation of geodesics (\ref{t}) and (\ref{phi}) have the
straightforward first integrals
\begin{equation} \label{t1}
\dot{t} = c_{1}\big(\frac{r}{(r-2M)}\big),
\end{equation}
\begin{equation} \label{phi1}
\dot{\phi} =\frac{c_{2}}{r(r-{\bf{\alpha}})}.
\end{equation}
The integration constants $c_{1}$ and $c_{2}$  can be found, if we know
the initial conditions $\gamma(\tau_{0})$ and
$\frac{d\gamma}{d\tau}(\tau_{0})$ for some $\tau_{0}\in \Re$. In the
case of Schwarzschild and Reissner-Nordstr\"om space-times, Clarke
\cite{C79} and also Wald \cite{W84} demonstrate that $c_{1}$
represents the total energy, $E$, per unit rest mass of a particle
as measured by a static observer, and $c_{2}$ represents the angular
momentum, $L$, per unit mass of a particle (see also \cite{MTW73}).
We recognize the constant $\varepsilon$ to represent the rest energy
per unit mass for massive particles (timelike curve, $\varepsilon
=-1$) or the rest energy for massless particles (null curves,
$\varepsilon = 0$), traveling along the given geodesic \cite{MTW73,
ABS75}.
\par
Using equations (\ref{t1}), (\ref{phi1}) and $\theta =
\frac{\pi}{2}$ in the constraint equation (\ref{Cr}), we have
\begin{equation}\label{8}
\dot{r}^2= E^2 -(1-\frac{2M}{r})\big[-\varepsilon +
\frac{L^2}{r(r-{\bf{\alpha}})}\big].
\end{equation}
Notice that we have not used equation (\ref{r}) for the following
reason: If we substitute $\dot{r}$ and its derivative with respect
to $\tau$ from the constraint equation (\ref{Cr}) and using
equations (\ref{t1}) and (\ref{phi1}) in equation (\ref{r}), then it is satisfied identically.
\par
Equation (\ref{8}) can be written as
\begin{equation}\label{9}
\dot{r}^2 = E^2 - V^2(r),
\end{equation}
where $V(r)$ is the "effective potential" defined by
\begin{equation}\label{V}
V^2(r) = (1-\frac{2M}{r})\big[-\varepsilon +
\frac{L^2}{r(r-{\bf{\alpha}})}\big],
\end{equation}
Equation (\ref{9}) is in the form of the equation of a
one-dimensional problem for a particle in a potential field $V(r)$.
\par
Since the left side of equation (\ref{8}) is positive or zero, the
energy $E$ of the trajectory must not be less than the potential
$V$. So for an orbit of a given $E$, the radial range is restricted
to those radii for which $V$ is smaller than $E$.
\par
In the following we  investigate the circular photon orbits in
the stringy charged black hole by analyzing the properties of the
effective potential.\\
In the case of the photon trajectories, putting $\varepsilon =0$,
the effective potential (\ref{V}) takes the form
\begin{equation}\label{VV}
V^2(r) = (1-\frac{2M}{r})\big[ \frac{L^2}{r(r-{\bf{\alpha}})}\big].
\end{equation}
Differentiating equation (\ref{9}) with respect to $\tau$ gives
$$
2\big(\frac{dr}{d\tau}\big)\big(\frac{d^2r}{d\tau^2}\big)=
-\frac{dV^2(r)}{dr}\frac{dr}{d\tau},
$$
or
\begin{equation}\label{10}
\frac{d^2r}{d\tau^2} =-\frac{1}{2}\frac{d}{dr}(V^2(r)).
\end{equation}
It is clear from equation (\ref{10}) that a circular orbit ($r$=
constant) is possible only at a minimum or maximum of $V^2(r)$. We
can quantitatively evaluate
$$
0= \frac{d}{dr}\Big[(1-\frac{2M}{r})\big[
\frac{L^2}{r(r-{\bf{\alpha}})}\big]\Big],
$$
and get
\begin{equation}\label{11}
r=\frac{1}{4}\Big((6M+{\bf{\alpha}})\pm
\sqrt{(6M+{\bf{\alpha}})^2-32M{\bf{\alpha}}}\Big).
\end{equation}
Since ${\bf{\alpha}}$ must be less than $r$, to keep the line
element (\ref{eq}) to be in Lorentzian metric, then for
${\bf{\alpha}}=2M$ and ${\bf{\alpha}}>2M$ no circular is possible.
From the relation (\ref{Q}) ${\bf{\alpha}}$ is  smaller than $M$
$({\bf{\alpha}}<M)$, therefor the larger of two roots given by
equation (\ref{11}) locates the maximum of the potential-energy
curve $V^2(r)$ defined by equation (\ref{VV}). Consequently, the
unstable circular orbit is always at the same radius
$r=\frac{1}{4}\Big((6M+{\bf{\alpha}})+
\sqrt{(6M+{\bf{\alpha}})^2-32M{\bf{\alpha}}}\Big)$, regardless of
$L$. For the smaller root no circular is possible.
\par
There are many possible ways to compare the effective potential for a massless particle in stringy charged black hole with that in Schwarzschild black hole. Here we assume ${\bf{\alpha}} < M$ and fix the value of angular momentum $L$; the values of $r$ and ${\bf{\alpha}}$ are calculated to obtain the plot of the effective potential. The choice of ${\bf{\alpha}}<< M$ allows the two paths to be the same. (This comparison is shown in figures (1) - (5).)

\par
 We notice that, when ${\bf{\alpha}} =0$, that is the Schwarzschild
case, the radius is $r=3M$ which is the same radius obtained by
Schutz \cite{S85} in the Schwarzschild black hole.

 \setcounter{equation}{0}
\section{Reissner-Nordstr\"{o}m metric}
A well-known simple solution of the Einstein-Maxwell equation is the
Reissner-Nordstr\"{o}m solution. This solution represents a
non-rotating charged black hole. Discussions of the basic properties
of this solution can be found in many places including works by
Chandrasekhar \cite{C83}, and Hawking and Ellis \cite{HE73}. The
metric is defined on a four-dimensional manifold and its typically
written in the form
\begin{equation}\label{RNS}
ds^2=-(1-\frac{2m}{r}+\frac{e^2}{r^2})dt^2 +
(1-\frac{2m}{r}+\frac{e^2}{r^2})^{-1} dr^2 +r^2(d\theta^2
+\sin^2\theta d\phi^2),
\end{equation}
where $m$ represents the gravitational mass and $e$ the electric
charge of the body.
\par
The equation of geodesics and the constraint of geodesic for the
line element (\ref{RNS}) are given as
\begin{equation}\label{tRNS}
\ddot{t}+\frac{2(-m+\frac{e^2}{r})}{r^2-2mr+e^2}\dot{r}\dot{t}=0,
\end{equation}
\begin{equation}\label{thetaRNS}
\ddot{\theta} +\frac{2}{r}\dot{r}\dot{\theta}
-\sin\theta\cos\theta\dot{\phi}^2=0,
\end{equation}
\begin{equation}\label{phiRNS}
\ddot{\phi} +\frac{2}{r}\dot{r}\dot{\phi}
+2\cot\theta\dot{\theta}\dot{\phi}=0,
\end{equation}
\begin{equation}\label{CrRNS}
-(1-\frac{2m}{r}+\frac{e^2}{r^2})\dot{t}^2 +
(1-\frac{2m}{r}+\frac{e^2}{r^2})^{-1} \dot{r}^2 +r^2(\dot{\theta}^2
+\sin^2\theta \dot{\phi}^2)=\varepsilon.
\end{equation}
We will assume, as section 2, that the orbit is in the $\theta =
\frac{\pi}{2}$ plane. Equation (\ref{thetaRNS}) shows that if
$\theta = \frac{\pi}{2}$  and $\dot{\theta}=0$ initially, then
$\ddot{\theta} = 0$ and the orbit remains in this plane.
\par
Equations (\ref{tRNS}) and (\ref{phiRNS}) can be integrated
directly, giving
\begin{equation} \label{t1RNS}
\dot{t} =\frac{ c_{3}r^2}{r^2-2mr+e^2},
\end{equation}
\begin{equation} \label{phi1RNS}
\dot{\phi} =\frac{c_{4}}{r^2},
\end{equation}
where the integrating constant $c_{3}$ represents the energy,
$\bar{E}$, (at $r \rightarrow \infty$) of a test particle and
$c_{4}$ the angular momentum, $\bar{L}$.
\par
Substituting (\ref{t1RNS}) and (\ref{phi1RNS}) in equation
(\ref{CrRNS}) and using the condition $\dot{\theta}=0$, we get
\begin{equation}\label{8RNS}
\dot{r}^2= \bar{E}^2 -(1-\frac{2m}{r} +
\frac{e^2}{r^2})\big[-\varepsilon + \frac{\bar{L}^2}{r^2}\big].
\end{equation}
Equation (\ref{8RNS}) can be written as
\begin{equation}\label{9RNS}
\dot{r}^2 = \bar{E}^2 - \bar{V}^2(r),
\end{equation}
where $\bar{V}(r)$ is the "effective potential" defined by
\begin{equation}\label{VRNS}
\bar{V}^2(r) = (1-\frac{2m}{r} + \frac{e^2}{r^2})\big[-\varepsilon +
\frac{\bar{L}^2}{r^2}\big].
\end{equation}
In this paper we restrict our tension to the photon trajectories,
by putting $\varepsilon =0$ in equation (\ref{VRNS}), the effective
potential becomes
\begin{equation}\label{VRNSP}
\bar{V}^2(r) = (1-\frac{2m}{r} + \frac{e^2}{r^2})\big[
\frac{\bar{L}^2}{r^2}\big].
\end{equation}
Differentiating (\ref{9RNS}) with respect to $\tau$, as in the
previous section, we get
\begin{equation}\label{10RNS}
\frac{d^2r}{d\tau^2} =-\frac{1}{2}\frac{d}{dr}(\bar{V}^2(r)).
\end{equation}
This equation shows that a circular orbit ($r$= const.) is possible
only at a minimum or maximum of $\bar{V}^2(r)$. We can quantitative
by evaluating
$$
0= \frac{d}{dr}\Big[(1-\frac{2m}{r} + \frac{e^2}{r^2})\big[
\frac{\bar{L}^2}{r^2}\big]\Big],
$$
which gives
\begin{equation}\label{RNS11}
r=\frac{3m}{2}\Big[1\pm \sqrt{1-\frac{8e^2}{9m^2}}\Big].
\end{equation}
This equation shows that the two radii are identical for
$\frac{e}{m}=\frac{3\sqrt{2}}{4}$ (in this case a minimum radius for
photon $r_{MIN} =\frac{3m}{2}$) and do not exist at all for
$\frac{e}{m}> \frac{3\sqrt{2}}{4}$. This indicates a qualitative
change in the shape of the curve of $\bar{V}^2$ for small
$\frac{e}{m}$. For $\frac{e}{m}< \frac{3\sqrt{2}}{4}$ the larger of
the two roots given by equation (\ref{RNS11}) locates the minimum of
the potential- energy curve $\bar{V}^2$ defined by equation
(\ref{VRNSP}), while the smaller root locates the maximum of the
potential-energy curve. Therefore, the circular orbit of the larger
radius will be stable in  contrast to circular orbit of the
smaller radius which will be unstable.

 \setcounter{equation}{0}
\section{Geodesic Deviation}
In this section, we use the tidal forces between free test particles
falling in a gravitational field to investigate different
properties between the stringy charged and vacuum black holes.
\par
Consider a sphere of two non-interacting particles falling freely
towards the center of the Earth. Each particle moves on a straight
line, but nearer the Earth fall faster because the gravitational
attraction is stronger. This means that the sphere does not remain a
sphere but is distorted into an ellipsoid with the same volume. The
same effect occurs in a body falling towards a spherical object in
general relativity, but if the object is a black hole the effect
becomes infinite as the singularity is reached. Jacobi vector fields
provide the connection between the behavior of nearby particles and
curvature, via the equation of geodesic deviation (Jacobi equation)
\begin{equation}\label{4.1}
\frac{D^{2}\eta^{a}}{D\tau^2} + R^{a}_{bcd}v^{b}v^{c}\eta^{d} = 0,
\end{equation}
where $v^a$ are the components of the tangent vector to geodesic and
$\eta^a$ are the components of the connecting vector between two
neighboring geodesics.
\par
   In order to investigate in detail the behavior of Jacobi fields
we consider a congruence of timelike geodesics (path of particles)
with timelike unit tangent vector $v$ ($g(v,v) = - 1$). We define at
some point $q$ on the geodesic $\gamma(\tau)$ dual bases $e_{0}^a,
e_{1}^a, e_{2}^a, e_{3}^a$  and $e_{a}^0, e_{a}^1, e_{a}^2, e_{a}^3$
of the tangent space $T_{q}M$ and dual tangent $T^{\star}_{q}M$
respectively in the following way \cite{D92} : We choose $e_{0}^a$
to be $v^a$ and $e_{1}^a, e_{2}^a, e_{3}^a$ as unit spacelike
vectors, orthogonal to each other and to $v^a$. If we parallelly
propagate the basis along the timelike geodesic $\gamma(\tau)$ (that
is, $\frac{D}{D\tau}e^{a}_{\alpha} = 0, \, \alpha = 1, 2, 3$),
$e^{a}_{0}$ will remain equal $v^a$, and $e_{1}^{a}, e^{a}_{2},
e^{a}_{3}$  will remain to orthogonal to $v^a$ (see \cite{HE73} p.
80). The frame $e_{0}^a, e_{1}^a, e_{2}^a, e_{3}^a$ is called
"parallel transported" (PT) frame. The orthogonal connecting vector,
$\eta^a$, between two neighboring timelike geodesics may be
expressed as $\eta^a  = \eta^{\alpha} e_{\alpha}^a \, (\eta^0 =
e_{\alpha}^{0}\eta^{\alpha} = 0)$.
\par
The geodesic deviation vector $\eta^a$ satisfy the following equation
\begin{equation}\label{4.2}
\frac{D\eta^{\alpha}}{D\tau} = \eta^{\alpha}_{;a}v^{a},
\end{equation}
\begin{equation}\label{4.3}
\frac{D^{2}\eta^{\alpha}}{D\tau^2} +
\tilde{R}^{a}_{bdc}e^{\alpha}_{a}v^{b}v^{c}e^{d}_{\beta}\eta^{\beta}
= 0,
\end{equation}
where $\eta^{\alpha}$ are the space-like components of the
orthogonal connecting vector $\eta^{a}$ connecting two neighboring
particles in free fall; $\eta^0 = 0$. The tilde denotes components
in the PT frame and the components of the Riemann tensor
$\tilde{R}^{a}_{bdc}$ are given by
\begin{equation}\label{4.4}
\tilde{R}^{a}_{bdc} =
e_{e}^{a}e_{b}^{f}e_{c}^{g}e_{d}^{h}R^{e}_{fgh}.
\end{equation}
From (\ref{eq}) the frame $e^{a}_{b}$ in Reissner-Nordstr\"{o}m
metric is given by :
\begin{equation}\label{4.5}
\begin{array}{ccc}
e_{0}^{a}& = (1 - \frac{2m}{r})^{-\frac{1}{2}}(0, 0, 0, 1),\\
e_{1}^{a}& = (1 - \frac{2m}{r} )^{\frac{1}{2}}(1, 0, 0, 0),\\
e^{a}_{2} &= \frac{(1-\frac{{\bf{\alpha}}}{r})^{-\frac{1}{2}}}{r} (0, 1, 0, 0),\\
e^{a}_{3} &=
\frac{(1-\frac{{\bf{\alpha}}}{r})^{-\frac{1}{2}}}{r\sin\theta}(0, 0,
1, 0).
\end{array}
\end{equation}
The components of $\eta^{\alpha}$ can be written as follows
$$
\eta^{\alpha} = (\eta^{1}, \eta^{2}, \eta^{3}) = (\eta^{r},
\eta^{\theta}, \eta^{\phi}).
$$
Using (\ref{4.4}), (\ref{4.5}), $v^{a} = e^{a}_{0}$ and the
components of Riemann tensor for the metric (\ref{eq}) (see
appendix), in (\ref{4.3}), we get
\begin{equation}\label{4.6}
\begin{array}{ccc}
\frac{D^{2}\eta^r}{D\tau^2} & =  \frac{2M}{r^3}\eta^r,\\
\frac{D^{2}\eta^{\theta}}{D\tau^2}& =  \frac{M({\bf{\alpha}} -2r)}
{2r^3(r-{\bf{\alpha}})}\eta^{\theta},\\
\frac{D^{2}\eta^{\phi}}{D\tau^2}& =  \frac{M({\bf{\alpha}}
-2r)}{2r^3(r-{\bf{\alpha}})}\eta^{\phi},
\end{array}
\end{equation}

In order to write equation (\ref{4.6}) in terms of ordinary
derivative, we must evaluate the second covariant derivative
$\frac{D^2}{D\tau^2}$. Using $e^{a}_{0} = v^{a}$, equation
(\ref{4.2}) takes the form

\begin{equation}\label{4.7}\frac{D\eta^{\alpha}}{D\tau} = \frac{d\eta^{\alpha}}{d\tau} +
\tilde{\Gamma}^{\alpha}_{ab}\eta^{b}v^{a},
\end{equation}

where $\tilde{\Gamma}^{\alpha}_{ab} = e_{e}^{\alpha}e_{a}^{f}e_{b}^{g}\Gamma^{e}_{fg}$.\\

Differentiating (\ref{4.7}) covariantly and using the Christoffel

components of metric (\ref{eq}) (see appendix), we can write

(\ref{4.6}) in the form

\begin{equation}\label{e1}
\frac{d^{2}\eta^r}{d\tau^2}  =  \frac{2M}{r^3}\eta^r,
\end{equation}
\begin{equation}\label{e2}
\frac{d^{2}\eta^{\theta}}{d\tau^2} = \frac{M({\bf{\alpha}}
-2r)}{2r^3(r-{\bf{\alpha}})}\eta^{\theta},
\end{equation}
\begin{equation}\label{e3}
\frac{d^{2}\eta^{\phi}}{d\tau^2} =  \frac{M({\bf{\alpha}}
-2r)}{2r^3(r-{\bf{\alpha}})}\eta^{\phi}.
\end{equation}

Equation (\ref{e1}) indicates that tidal force in radial direction will
stretch an observer falling in this fluid. To keep the line element
(\ref{eq}) to be in Lorentzian metric, $\alpha$ should be less than
$r$. Therefore equations (\ref{e2}) and (\ref{e3}) indicate a
pressure or compression in the transverse directions.

\section{ Solution of the equations  (\ref{e1})-(\ref{e3})}
To solve the equations  (\ref{e1})-(\ref{e3}), we consider the special case when freely falling particles have zero angular momentum ($L=0$). From equation (\ref{8}) we obtain the relation between the radial coordinate, $r$, and the affine parameter, $\tau$, in the following form
\begin{equation}
\frac{dr}{d\tau}=-\sqrt{(E^2-1)+\frac{2M}{r}}.
\end{equation}
Using this relation in equations  (\ref{e1})-(\ref{e3}), we consider, without loss of generality, the case when  $E^2=1$, we obtain
\begin{equation}\label{e1-}
\frac{d^{2}\eta^r}{dr^2} - \frac{1}{r}\frac{d\eta^r}{dr}=  \frac{1}{r^2}\eta^r,
\end{equation}
\begin{equation}\label{e2-}
\frac{d^{2}\eta^{\theta}}{dr^2} - \frac{1}{2r} \frac{d\eta^{\theta}}{dr}= - \frac{(2r-{\bf{\alpha}})}{4r^2(r-{\bf{\alpha}})}\eta^{\theta},
\end{equation}
\begin{equation}\label{e3-}
\frac{d^{2}\eta^{\phi}}{dr^2}- \frac{1}{r}\frac{d\eta^{\phi}}{dr} =  -\frac{(2r-{\bf{\alpha}})}{4r^2(r-{\bf{\alpha}})}\eta^{\phi}.
\end{equation}
Solving the above equations, using MAPLE, we obtain the space-like components of the geodesic deviation vector in the following form
\begin{equation}\label{e1--}
\eta^r = \frac{C_{1}}{\sqrt{r}}+C_{2}r^2,
\end{equation}
$$
\eta^{\theta} = C_{3}F_{1}\big(\frac{1}{2}+b, b, -\frac{1}{2}+2b, \frac{r}{{\bf{\alpha}}}\big) r^b(r-{\bf{\alpha}}) +
$$
\begin{equation}\label{e2--}
C_{4}F_{2}\big(\frac{3}{2}-b, 2-b, \frac{5}{2}-2b, \frac{r}{{\bf{\alpha}}}\big)r^{(\frac{3}{2}-b)}(r-{\bf{\alpha}}),
\end{equation}
$$
\eta^{\phi} = C_{5}F_{1}\big(\frac{1}{2}+b, b, -\frac{1}{2}+2b, \frac{r}{{\bf{\alpha}}}\big) r^b(r-{\bf{\alpha}}) +
$$
\begin{equation}\label{e3--}
 C_{6}F_{2}\big(\frac{3}{2}-b, 2-b, \frac{5}{2}-2b, \frac{r}{{\bf{\alpha}}}\big)r^{(\frac{3}{2}-b)}(r-{\bf{\alpha}}),
\end{equation}
where $C_1, C_2, C_3$ and $C_4$ are constants of integration and $b=\frac{3}{4} \pm \frac{\sqrt{5}}{4}$. $F_1$ and $F_2$ are hypergeometric functions.\\
The spacial solution given by  (\ref{e1--})-(\ref{e3--}) allows one, by a suitable choice of initial conditions and constants appearing in  (\ref{e1})-(\ref{e3}), to describe various physical situations connected with relative motion of freely falling particles.
\par
At the singularity $r=0$ the radial component $\eta^r$ of geodesic deviation vector becomes infinite, while the transverse components $\eta^{\theta}$ and $\eta^{\phi}$  vanish. Therefore, the behavior of geodesic deviation vector in the space-time under consideration is the same as in the case of Schwarzschild space-time, when the singularity, $r=0$, is approached. In fact the Schwarzschild singularity has time-like geodesic for which one component of the geodesic deviation vector becomes infinite while the other two vanish \cite{MTW73}. The singularity $r=0$ is a strong curvature singularity\footnote{A singular point is called a strong curvature singularity if any object hitting it is crushed to zero volume \cite{T77}.}, as the Schwarzschild singularity \cite{T77}, for the two vanishing components of geodesic deviation vector go to zero faster than the remaining component becomes infinite. The misbehavior of these components could in concert so that it is not visible in the volume element magnitude.
\par
At the singularity $r= {\bf{\alpha}}$, the component  $\eta^r$ has a finite value, while the other two components, $\eta^{\theta}$ and $\eta^{\phi}$, vanish. Then the volume element, defined by the three space-like components of geodesic deviation vector \cite{P72}, \cite{T77}, along a time-like geodesic vanishes as the geodesic approaches  this singularity.

\section{Conclusion}
In this paper we have studied the circular photon orbits in charged
black holes by analyzing the properties of effective potential.
Considering the light-like geodesics, we classified and analyzed the
different cases between the stringy charged black hole and the
vacuum solutions. These differences arose from considering the orbits
associated with stable and unstable circular orbits. In the context
of the Schwarzschild geometry there is an unstable circular orbit
which is always at the same radius, $r=3M$. In the
Reissner-Nordstr\"{o}m space-time  there are two radii, the circular
orbit of the larger radius will be stable while that of the smaller
radius will be unstable. In the case of stringy charged black hole
there is an unstable circular orbit as the Schwarzschild black hole,
but the difference is that the radius in the case of stringy charged
black hole depends on the charge $Q$.
\par
Equations (\ref{e1})-(\ref{e3}) provide the explicit expressions of
the relative accelerations in a stringy charged black hole. Two
comments are worth making about expression (\ref{e1}). First there
is no divergence in the radial direction at $r=2m$. Secondly, the
tidal field at the horizon in radial direction is larger for smaller
black hole. This is simply because
$$
\frac{d^{2}\eta^r}{d\tau^2}\sim \frac{2M}{r^3}\eta^r\sim
\frac{1}{M^2}\eta^r \quad at\,\,\, r\sim M.
$$
\par
The radial component of the geodesic deviation vector field,
equation (\ref{e1}), is the same as the component obtained in the
case of Schwarzschild black hole and different from the component
obtained in the case of Reissner-Norstr\"{o}m. Consequently, in the
case of stringy charged and Schwarzschild black holes, the tidal
forces in the radial direction will stretch an observer falling in
these black holes, while in the case of Reissner-Norstr\"{o}m the
radial component depend on the quantity $2e^2-2mr$ to indicate a
tension or stretching in the radial direction. In the transverse
directions the relative acceleration components (see equations
(\ref{e2}) and (\ref{e3})) indicate a compression in these
directions. This property is similar as in the case of Schwarzschild
black hole, while in the case of Reissner-Nordstr\"{o}m depends on
the quantity $e^2-2mr$ to indicate compression or tension in these
directions.
\par
To describe various physical situations connected with relative
motion of freely falling photons, in stringy charged black hole,
we tried to solve  equations  (\ref{e1})-(\ref{e3}) by considering a special case. We found that any object hitting the singularity $r={\bf{\alpha}}$ is crushed to zero volume.\\
 \setcounter{equation}{0}
\section*{Appendix}

We use $(x^0, x^1,x^2,x^3)= (t, r, \theta, \phi)$ so that the
non-vanishing Christoffel symbols of the second kind of the line
element (\ref{eq}))are
$$
\begin{array}{cccc}
\Gamma^{1}_{11}& = \frac{M}{r(2M-r)},&
\Gamma^{2}_{12} & = \frac{\alpha -2r}{2r(\alpha -r)},\\
\Gamma^{1}_{22} & =  \frac{(r-2M)(\alpha -2r}{2r}, & \Gamma^{3}_{13} &=\frac{\alpha -2r}{2r(\alpha -r)},  \\
\Gamma^{1}_{33} & = \frac{(r-2M)(\alpha -2r)}{2r}\sin^2\theta, & \Gamma^{2}_{33} & = -\sin\theta\cos\theta ,  \\
\Gamma^{1}_{00}  &=  \frac{M(r-2M)}{r^3}, & \Gamma^{3}_{23}& =\cot\theta  ,  \\
 \Gamma^{0}_{10}&  = - \frac{M}{r(r-2M)}.
 \end{array}
$$
 The non-zero Riemann tensor are:
$$
\begin{array}{ccc}
R^{1}_{212} & =  \frac{2M(r-\alpha)(2r-\alpha) +\alpha^2(2M-r)}{4r^2(\alpha -r)},\\
  R^{1}_{313} & =  \frac{2M(r-\alpha)(2r-\alpha) +\alpha^2(2M-r)}{4r^2(\alpha -r)}\sin^2\theta,  \\
R^{1}_{010} & =  \frac{2M(2M-r)}{r^4},\\
  R^{2}_{323} & = \frac{8Mr(r-\alpha)-\alpha^2(r-2M)}{4r^2(r-\alpha)}\sin^2\theta,  \\
 R^{2}_{020} & = R^{3}_{030}=
 \frac{M(\alpha-2r)(2M-r)}{2r^4(r-\alpha)}.
\end{array}
$$


\begin{thebibliography}{99}
\bibitem{B75} J. D. Bekenstein, Ann. Phys. (N. Y.), {\bf{91}}, 75
(1975).
\bibitem{AG99} E. Ayon-Beato and A. Garcia, Gen. Relat. Grav.
{\bf{31}}, 629 (1999).
\bibitem{G91} D. Garfinkle, G. T. Horowitz and A. Strominger,
Phys. Rev. {\bf{D43}}, 3140 (1991); {\bf{D45}}, 3888(E) (1992).
\bibitem{M58} C. M{\o}ller, Ann. Phys. (N. Y.), {\bf{4}}, 347
(1958).
\bibitem{G04} R. M. Gad, Astrophys. Space Sci. {\bf{295}}, 459
(2004).
\bibitem{X} S. S. Xulu, Astrophys. Space Sci. {\bf{283}}, 23 (2003).
\bibitem{PSSTW91} J. Preskill, P. Schwarz, A. Shapere, S. Trivedi
and F. Wilczek, Mod. Phys. Lett., {\bf{6}}, 2353 (1991).
\bibitem{HS} J. A. Harvey and A. Strominger, "Quantum aspects of
black holes", Preprint EFI-92-41, hep-th/9209055.
\bibitem{FAE} F. A. E. Piran, Phys. Rev. {\bf{105}}, 1089 (1957).
\bibitem{T83} K. P. Tod, Proc. R. Soc. Lond. {\bf{A388}}, 467
(1983).

\bibitem{MTW73} C. Misner, K. Thorne and J. Wheeler, (1973),
"Gravitation", Freeman, San Francisco.
\bibitem{ABS75}R. Adler, M. Bazin and M. Schiffer, (1975),
"Introduction to General Relativity", (McGrow-Hill, New York, 2nd.
ed.).
\bibitem{S85}B. F. Schutz, (1985), " A First Course in
General Relativity" ( Cambridge Uni. Press, Cambridge, London, New
York, New Rochelle, Melbourne Sydney).
\bibitem{C83} S. Chandrasekhar, (1983), "The Mathematical Theory
of Black Holes", (Oxford Uni. Press, Cambridge, England).
\bibitem {HE73} S. W. Hawking and G. F. R. Ellis, "The Larger
Scale Structure of Space-time", (Cambridge Uni. Press, Cambridige).
\bibitem{D92} R. D'Invermo, "Introducing Einstein's Relativity",
Oxford University Press, New York, (1992).
\bibitem{AG05} M. Abdel-Megied and R. M. Gad, Chaos, Solitons and
Fractals, {\bf{23}}, 313 (2005).
\bibitem{KMMH01} R. Kerner, J. Martin, S. Mignemi and J. E. Van Holten, Phys. Rev. D {\bf{63}}, 027502 (2001).
\bibitem{EE97} G. F. R. Ellis and H. van Elst, "Deviation of geodesics in FLRW spacetime geometries", arXiv: gr-qc/9709060.
\bibitem{KK03} R. Koley, S. Pal and S. Kar, Am. j. Phys., {\bf{71}}, 1037 (2003)
\bibitem{GK09} S. Ghosh and Sayan Kar,"Geodesics and geodesic deviation in warped spacetime with a time dependent extra dimension",{\bf{ arXiv:0904.2321v1[gr-qc]}}.
\bibitem {STW91} A. Shapere, S. Trivedi and F. Wilczek, Mod. Phys.
Lett. {\bf{A6}}, 2677 (1991).

\bibitem{C79} C. J. S. Clark, (1979), "Elementary General
relativity",(Edwaed Arnold, London).
\bibitem{W84} R. M. Wald, (1984), "General Relativity",
(Chicago and London).
\bibitem{P72} R. Penrose, (1972), "Techniques of Differential Topology in Relativity", (SIAM, Philadelphia, p. 60).
\bibitem{T77} F. J. Tipler, Phys. Lett. {\bf{64A}}, 8 (1977).
\end{thebibliography}
\end{document}